\RequirePackage{plautopatch} 
\documentclass[12pt]{article}
\usepackage[utf8]{inputenc}

\usepackage[dvipdfmx]{graphicx}
\usepackage{url}
\usepackage{listings,jvlisting}
\usepackage{amssymb}
\usepackage{tabularx}
\usepackage{amsmath}
\usepackage{mdframed}
\usepackage{graphicx}
\usepackage{setspace}
\usepackage{here}
\usepackage{authblk}
\usepackage{mathtools}
\usepackage{amsthm}
\usepackage{algorithmic}
\usepackage{algorithm}
\usepackage{arydshln}

\def\ga{{\gamma}}

\def\bbe{{\text{\boldmath $\beta$}}}

\def\bbeh{{\hat \bbe}}

\def\gah{{\hat \ga}}

\def\muh{{\widehat \mu}}

\def\bbeh{{\widehat \bbe}}

\def\U{{\text{\boldmath $U$}}}

\def\Z{{\text{\boldmath $Z$}}}

\def\vh{{\hat v}}

\def\Sh{{\widehat S}}

\begin{document}
\title{Comparison of Baseline Covariate Adjustment Methods for Restricted Mean Survival Time}
\author[1]{Keisuke Hanada}
\author[1]{Junji Moriya}
\author[1,2]{Masahiro Kojima\footnote{Address: Biometrics Department, R\&D Division, Kyowa Kirin Co., Ltd.
Otemachi Financial City Grand Cube, 1-9-2 Otemachi, Chiyoda-ku, Tokyo, 100-004, Japan. Tel: +81-3-5205-7200 \quad
E-Mail: masahiro.kojima.tk@kyowakirin.com}}
\affil[1]{Kyowa Kirin Co., Ltd}
\affil[2]{The Institute of Statistical Mathematics}
\maketitle

\abstract{\noindent
The restricted mean survival time is a clinically easy-to-interpret measure that does not require any assumption of proportional hazards. We focus on two ways to directly model the survival time and adjust the covariates. One is to calculate the pseudo-survival time for each subject using leave-one-out, and then perform a model analysis using all pseudo-values to adjust for covariates. The pseudo-survival time is used to reflect information of censored subjects in the model analysis. The other method adjusts for covariates using subjects for whom the time-to-event was observed while adjusting for the censored subjects using the inverse probability of censoring weighting (IPCW). This paper evaluates the performance of these two methods in terms of the power to detect group differences through a simple example dataset and computer simulations. The simple example illustrates the intuitive behavior of the two methods. With the method using pseudo-survival times, it is difficult to interpret the pseudo-values. We confirm that the pseudo-survival times are different from the actual data obtained in a primary biliary cholangitis clinical trial because of the many censored data. In the simulations, the method using IPCW is found to be more powerful. Even in the case of group differences with respect to the censor incidence rates and covariates, the method using IPCW maintains a nominal significance level for the type-1 error rate. We conclude that the IPCW method should be used to estimate the restricted mean survival time when adjusting the covariates.

}
\par\vspace{4mm}

\section{Introduction}
\label{sec1}
The proportional hazards (PH) Cox model is a well-known method for adjusting the baseline covariates when using time-to-event data. The PH Cox model evaluates the influence of baseline covariate parameters through hazard ratios. However, a statistically valid assessment of the PH Cox model requires an assumption of PH. In fact, in anticancer drugs such as immune checkpoint inhibitors, the survival curves of the treatment and control groups coincide for some time after the start of treatment, and then separate~\cite{borghaei2015,robert2015}. For such survival curves, the PH assumption may not hold. Spruance et al.~\cite{spruance2004} reported that physicians have difficulty in helping patients to understand the treatment effect when assessed in terms of hazard ratios. Therefore, in recent years, the restricted mean survival time (RMST), which does not require the PH assumption and uses familiar averages, has attracted considerable attention. Because the RMST uses the mean survival time as a summary measure, differences between treatment groups are easy to interpret. The mean survival time is calculated using the area under the survival percentage curve, as estimated by the Kaplan-Meier method, which is adjusted for censored subjects. In other words, the mean survival time is calculated as $\muh=\int^\infty_0\Sh(t)dt$, where $\Sh(t)$ is the Kaplan-Meier estimator at time $t$. The Kaplan-Meier estimator, however, cannot be defined when the largest observation is censored. To avoid this, it is necessary to restrict time to a certain time point $\tau$.

A number of baseline covariate adjustment methods for RMST have been proposed. The adjustment methods fall into indirect and direct categories. The indirect method uses a PH Cox model with estimated cumulative hazards to derive the RMST~\cite{karrison1987,zucker1998}. Although the advantage of RMST is that it can be calculated without assuming PH, the indirect adjustment method requires the assumption of PH, which may not be appropriate. Additionally, it requires the estimation of cumulative hazards. To overcome these problems, direct methods, which model survival and adjust the covariates directly without the assumption of PH, have been developed by Andersen et al.~\cite{andersen2004} and Tian et al.~\cite{tian2014}. An RMST that is adjusted for time-dependent covariates after initial treatment has been proposed by Zhang et al. \cite{zhang2022}.

In the course of a controlled clinical trial with a primary endpoint that can be characterized as a time-to-event, we had planned to analyze the difference in RMSTs between two groups by adjusting the baseline covariates. However, we were faced with the question of whether Andersen's or Tian's method should be applied. Ambrogi et al.~\cite{ambrogi2022} examined the accuracy of parameter estimation for Andersen's and Tian's methods, but the difference in statistical power between these methods has not yet been confirmed. Karrison and Kocherginsky~\cite{karrison2018} examined the effects of adjustment for covariates, but did not identify the differences among adjustment methods.

In this research, we compare the performance of Andersen's and Tian's methods in terms of the power to detect group differences through a simple example dataset and simulations. We illustrate the intuitive behavior of the two methods in the simple example, and evaluate the detection power through computer simulations with Cox's model as a reference. Simulations are performed under various scenarios. Because Andersen's and Tian's methods can be easily performed with the RMSTREG PROCEDURE in SAS, we use the SAS environment to conduct the simulations. The SAS program code used in the simulations is summarized in the Supplemental Material. Additionally, we confirm the difference in RMSTs adjusted by covariates using data from two actual trials.

The remainder of this paper is organized as follows. Section 2 introduces Andersen's and Tian's methods. Section 3 describes the setting and results of the computer simulations, before Section 4 describes the results of the two actual clinical trials. We conclude this paper with a discussion of our results in Section 5. The program code used for the SAS simulations, as well as the SAS analysis program code for the two actual clinical trials, is included in the Supplemental Material. 

\section{Methods}\label{sec2}
This section first presents a definition of RMST. Next, we describe the two methods of adjustment for the baseline covariates, one using the pseudo-survival times (PSTs) and the other based on inverse probability censoring weighting (IPCW).

\subsection{Definition of RMST}
Suppose that $T$ is the survival time subject to right censoring by a censoring time $C$, which is independent of $T$.
The observable survival time is $R=\min(T,C)$ with the censoring indicator $I(T \le C)$.
For a sample size of $n$, the data of the $i$-th subject is $R_i$ for $i=1,2,\ldots, n$.
For any restricted time $\tau>0$, the RMST is defined as
\begin{align}
    \mu(\tau) &= E[\min(T, \tau)] = \int_{0}^{\tau} S(t) dt
\end{align}
where $S(t)=\Pr(T>t)$ is the survival function.
The estimator of RMST $\mu(\tau)$ is $\hat{\mu}(\tau)=E[\min(R,\tau)]=\int_0^{\tau} \hat{S}(t)dt$.
In a two-arm clinical trial with survival functions $S_0(t)$ and $S_1(t)$ in the control and treatment groups, respectively, the difference $d$ in RMST between the groups is given by
\begin{align}
    d 
    &= \int_0^{\tau} S_1(t) dt - \int_0^{\tau} S_0(t) dt
    = \int_0^{\tau} [S_1(t) - S_0(t)] dt
\end{align}
i.e., $d$ is the area between the survival curves.
In the standard case, we can test the null hypothesis $d=0$ by comparing $z=\hat{d}/\mathrm{SE}(\hat{d})$ with a student-$t$ or (in large samples) normal reference distribution~\cite{royston2013}.

For the covariate-adjusted survival function, a method using Cox's model has been proposed~\cite{karrison1987,zucker1998}, but this is computationally complex because the nonparametric hazard function must be calculated. Therefore, to resolve the problems of the indirect method, direct methods for modeling survival and adjusting covariates without the assumption of PH were developed by Andersen et al.~\cite{andersen2004} and Tian et al.~\cite{tian2014}.

\subsection{Model-based estimation using PSTs}
Andersen et al.~\cite{andersen2004} proposed a covariate adjustment method using PSTs to consider the censored subjects.
Let $X_i=$min$(R_i,\tau)$, $i=1, \dots, n$, be random variables. We assume that the RMST is $\mu(\tau) = E[X_i]$ and the estimator of $\mu(\tau)$ is $\muh(\tau)$. The PST of the $i$-th subject is defined as
\begin{align}
    \muh_i &= n \muh(\tau) - (n-1) \muh^{-i}(\tau)
\end{align}
where $\muh(\tau)$ is calculated as the RMST using all subjects and $\muh^{-i}(\tau)$ is the ``leave-one-out" estimator for $\mu(\tau)$ based on $X_j, j \ne i$.

With $\muh_i$ as the objective variable, regression analysis can be performed with a generalized linear model for the $q$ covariates $\Z_i$. The generalized linear model is defined as follows:
\begin{align}
\label{eq:glm}
    g(\muh_i) &= \bbe^T \Z_i
\end{align}
with the link function $g(\cdot)$, where $\bbe$ is a $q$-dimensional parameter vector. For the survival time, the link function is set to the identity function or the log function.
The regression parameters $\bbe$ for PST $\muh_i$ can be estimated using generalized estimating equations as
\begin{align}
\label{eq:gee}
    \U_{PST}(\bbe) 
    &= \sum_{i=1}^n \U_i(\bbe) \nonumber \\
    &= \sum_{i=1}^n \left( \{\muh_i - g^{-1}(\bbe^T \Z_i)\}\frac{\partial}{\partial \bbe} g^{-1}(\bbe^T \Z_i)  \right) \nonumber \\
    &= 0.
\end{align}
We state the generalized estimating equation in the appendix for the cases in which the identity and log functions are applied as the link function.

Andersen et al. \cite{andersen2003} showed that consistent estimates of $\bbe$ could be obtained from \eqref{eq:gee} and that variance estimates for the solution $\bbeh$ could be obtained from the standard sandwich estimator
\begin{align}
    \hat{V} &= I(\hat{\bbe})^{-1} \hat{\mbox{var}}\{U_{PV}(\bbe)\} I(\hat{\bbe})^{-1},
\end{align}
with 
\begin{align}
    &I(\bbe) = \frac{1}{\vh^2}\sum_{i=1}^n \left( \frac{\partial g^{-1}(\bbe^T \Z_i)}{\partial \bbe} \right)^T \left( \frac{\partial g^{-1}(\bbe^T \Z_i)}{\partial \bbe} \right) \\
    &\hat{\mbox{var}}\{U_{PV}(\bbe)\} = \frac{1}{\vh^4}\sum_{i=1}^n U_i(\hat{\bbe})U_i(\hat{\bbe})^T,
\end{align}
where $\vh^2$ is the variance of $\muh_i$.

\subsection{Model-based estimation using IPCW}
Tian et al.~\cite{tian2014} proposed a covariate adjustment method weighted by the survival function of censoring. The influence of the censored data can be adjusted by weighting the inverse of the survival function based on IPCW. 
The IPCW estimating function for $\bbe$ in model \eqref{eq:glm} is
\begin{align}
    U_{IPCW}(\bbe) &= \sum_{i=1}^n \frac{I(X_i\le C_i) \{ X_i - g^{-1}(\bbe^T \Z_i)\}}{\hat{S}(X_i)} \Z_i=0,
\end{align}
where $\hat{S}(\cdot)$ is the Kaplan-Meier estimator of the censoring time $C$ based on $R_i$ and the censoring indicator $I(X_i\le C_i)$ for $i=1, \dots, n$.
We state the generalized estimating equation in the appendix for the cases in which the identity and log functions are applied as the link function.

Let $\hat{\bbe}$ be the unique root of $U_{IPCW}(\bbe)=0$.
An important property for building a prediction model is that $\hat{\bbe}$ converges to a constant $\bar{\bbe}$ in probability, even when model \eqref{eq:glm} is misspecified.
Additionally, $n^{1/2}(\hat{\bbe}-\bar{\bbe})$ converges weakly to a normal distribution with mean zero as $n \to \infty$~\cite{tian2014}.

\subsection{Examples of Andersen's and Tian's methods}
We now illustrate the intuitive behavior of Andersen's and Tian's methods with a simple example dataset. The example dataset is presented in Table \ref{MK_Table3}. This dataset assumes that 12 subjects are followed and their survival times observed, and consists of the survival time (weeks), censoring (yes/no), and age (covariate). The cut-off is assumed to be 100 weeks. We add the PSTs derived by leave-one-out in Andersen's method. The PSTs are almost identical to the original survival times when the event occurs. However, when the data are censored, the time at which the event is supposed to have occurred is calculated. The link function is set to identity to understand the interpretation of the analysis results easily.

\begin{table}[H]
  \begin{center}
\caption{Example dataset\label{MK_Table3}}
\begin{tabular}{|c|c|c|c|c|c|c|c|c|c|}\hline
\multicolumn{5}{|c|}{Treatment group}& \multicolumn{5}{c|}{Control group}\\\hline
ID&ST & Censor & Age & PST &ID& ST & Censor & Age & PST\\ \hline
1&20 & Yes & 60 & 78.4 &7& 20 & No & 70 & 20.0\\
2&40 & No & 80 & 30.4 &8& 30 & Yes & 60 & 78.4\\
3&60 & Yes & 70 & 100.4 &9& 40 & No & 60 & 30.4\\
4&80 & No & 70 & 75.4 &10& 50 & No & 80 & 42.9\\
5&100 & Yes & 60 & 106.6 &11& 80 & Yes & 70 & 106.6\\
6&100 & Yes & 60 & 106.6 &12& 100 & Yes & 60 & 106.6\\\hline
\end{tabular}
\\
      \footnotesize{ST: Survival Time, PST: Pseudo-Survival Time}
  \end{center}
\end{table}

\begin{figure}[H]
  \begin{center}
  \includegraphics[width=15cm]{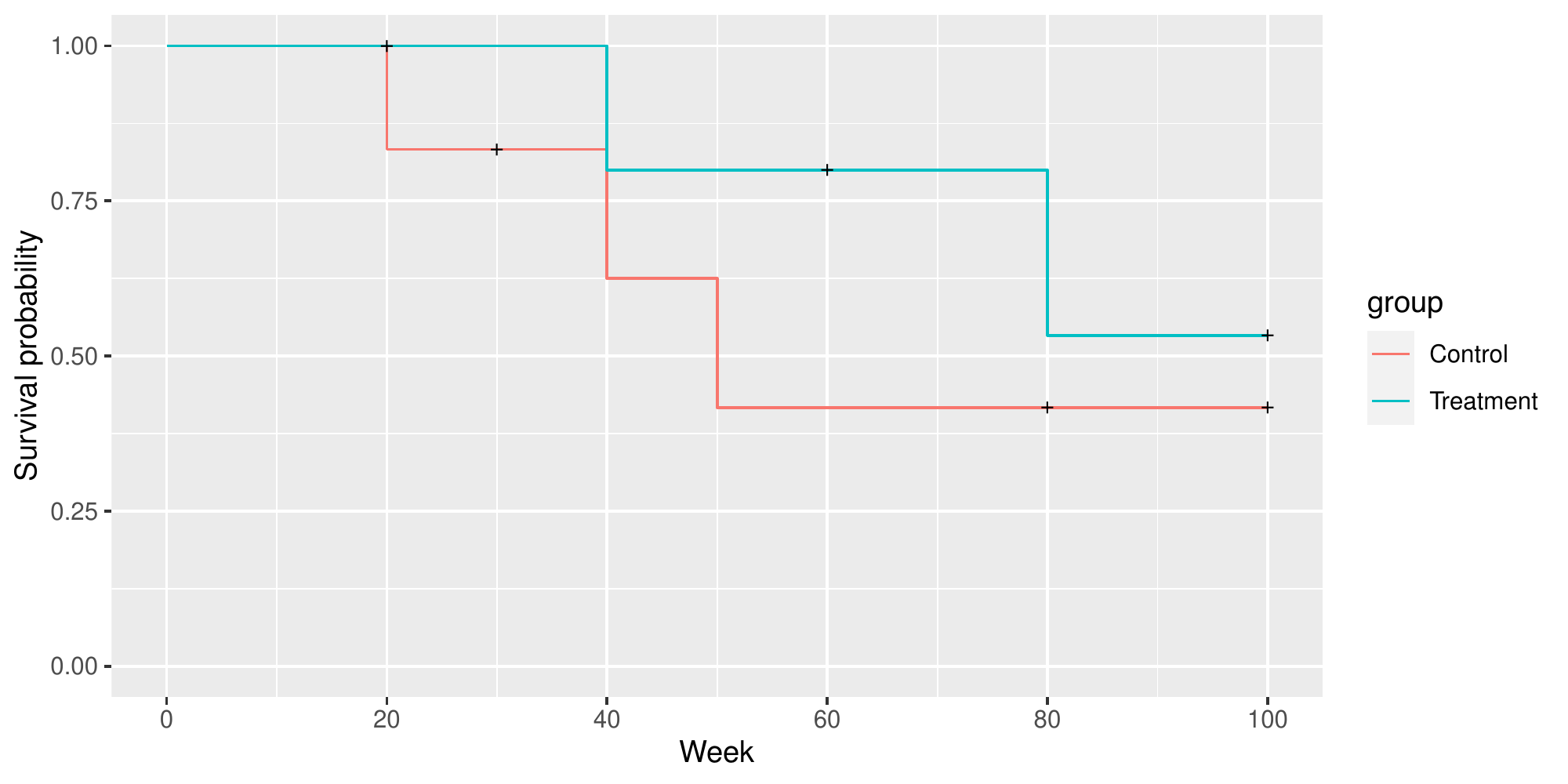}
  \caption{Survival curve of original survival data of example dataset}
  \label{MK_Figure5}
      \footnotesize{+ means censored.}
  \end{center}
\end{figure}

\begin{figure}[H]
  \begin{center}
  \includegraphics[width=15cm]{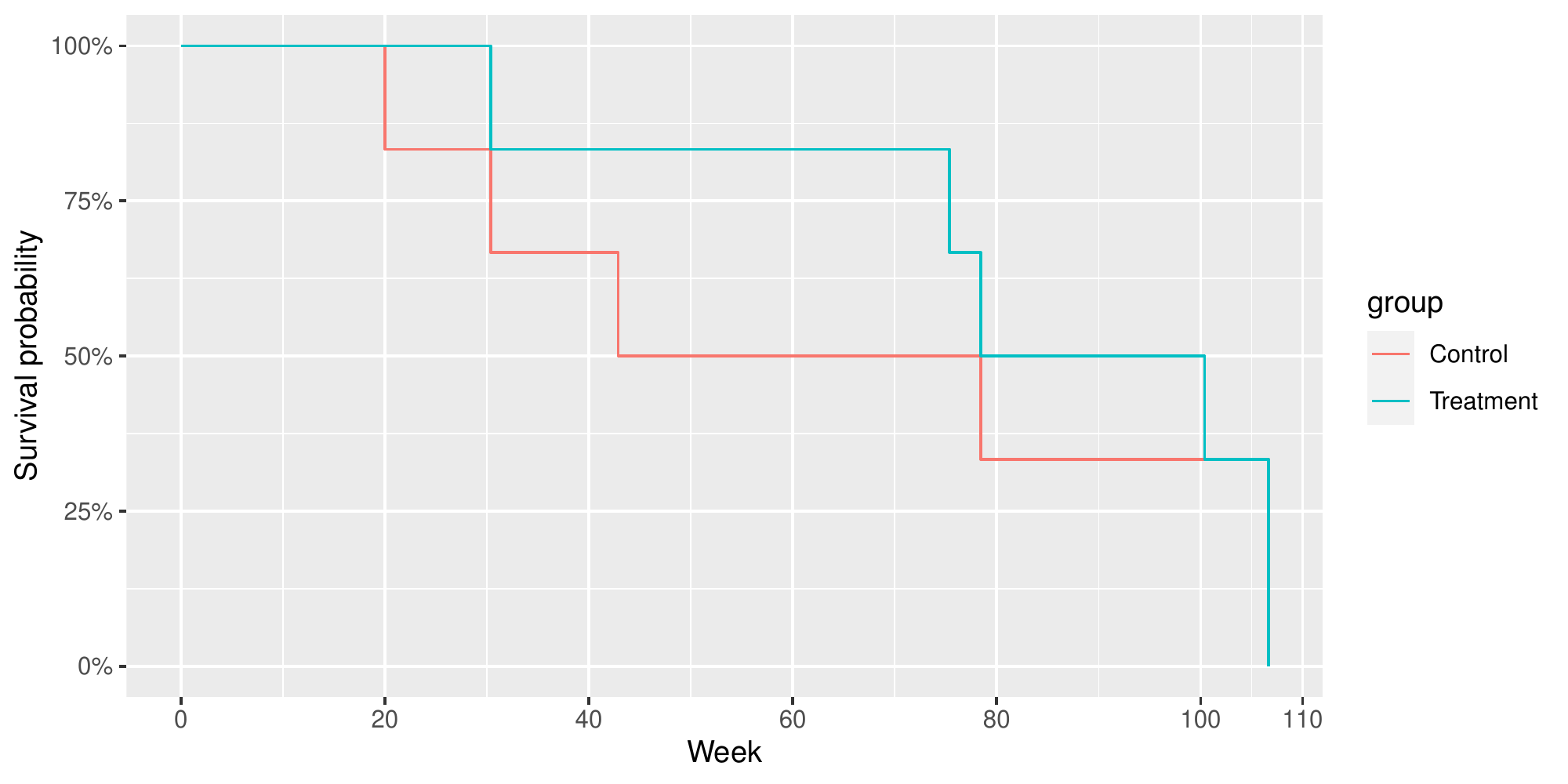}
  \caption{Survival curve of pseudo-survival data of example dataset}
  \label{MK_Figure6}
  \end{center}
\end{figure}

The survival curves of the original data are shown in Figure \ref{MK_Figure5}, and the survival curves of the PST are presented in Figure \ref{MK_Figure6}. From these figures, note that the mean survival time for both groups has increased through the use of PST.

The difference in RMSTs between the two groups (treatment group versus control group) until the cut-off point of $100$ weeks is summarized in Table \ref{MK_Table4}. The difference in the two RMSTs does not change under Andersen's method, even if age (covariate) is included, because the average age is equal in both groups when censored patients are included. However, the average age of patients for whom the event was observed is 75 (years) in the actual group and 70 (years) in the placebo group. Tian's method reduces the difference in the two RMSTs by adjusting the covariate. The difference between the groups can be reduced by considering the censored information compared with that in the naive RMSTs.

\begin{table}[H]
  \begin{center}
\caption{Differences in RMSTs in the example dataset\label{MK_Table4}}
\begin{tabular}{|c|c|c|c|c|c|}\hline
Parameter & Naive & Andersen & Tian & Andersen (age) & Tian (age)\\\hline
Diff & 29.1 & 18.8 & 25.0 & 18.8 & 24.5\\
Age & - & - & - & -2.1 & -2.3\\\hline
\end{tabular}
\\
      \footnotesize{Andersen/Tian: statistical model includes only the treatment group as a fixed effect. Andersen (age)/Tian (age): statistical model includes the treatment group as a fixed effect and age as a covariate.}
  \end{center}
\end{table}

We also consider two additional cases in which the age data are changed to confirm the influence caused by age. In the first case, there is no difference in average age between groups for all patients, including those censored, but there is a large difference in average age between groups including only patients for whom the event occurs. In the second case, there is a difference in average age for all patients, including those censored. Details of additional analyses are provided in the appendix. For the first case, Andersen's method does not change the RMST difference results, while Tian's method increases the difference. In the second case, Andersen's method reduces the difference, whereas Tian's method does not change the difference.

The example data and analysis program code are presented in the Supplemental Material.

\section{Simulation Study}
We now present the results of simulations to evaluate the methods for deriving the RMST with covariate adjustments.

\subsection{Simulation setup}
We consider 16 simulation scenarios to evaluate the performance of Andersen's and Tian's methods in terms of the statistical power for identifying the difference in RMSTs between treatment and control groups. Eight scenarios are based on the case where the PH assumption between the two groups holds. In scenario 1 (PH-S1), the true survival curves are consistent across the two groups, allowing us to confirm the type-1 error rate. In PH-S2, the percentage of censored data in the treatment group is higher than in the control group. This enables us to confirm the performance of Andersen's and Tian's methods, which handle censored data differently. In PH-S3, the survival curve of the treatment group is better than in the placebo group. In PH-S4, the percentage of censored data in the treatment group is higher than in the control group. Even in randomized trials, the covariates may differ between the two groups. Thus, in PH-S5--PH-S8, the averages of the true covariates differ between the two groups, allowing us to confirm that the covariates can be adjusted when differences occur. The remaining eight scenarios consider settings in which the slope of the survival function changes at a certain time, called a change point (CP). In CP-S1--CP-S4, the two survival functions intersect. In CP-S5--CP-S8, the control group has many events up to the CP, and few events after the CP. 

For PH-S1--PH-S8, we use the regression model below to generate the simulation data. Let the survival times $T$ be given by
\begin{align}
    \log T = \beta_0 + \beta_1 x_1 + \beta_2 x_2 + \log(\varepsilon),
\end{align}
where $\beta_0$ is the intercept parameter, $\beta_1$ is a parameter reflecting the difference between treatment ($x_1=1$) and control ($x_1=0)$, $\beta_2$ is a parameter reflecting the covariate, $x_2$ is normally distributed with variance $20$, and $\varepsilon$ is a random variable obeying a Weibull distribution. 

\begin{figure}[H]
  \begin{center}
  \includegraphics[width=15cm]{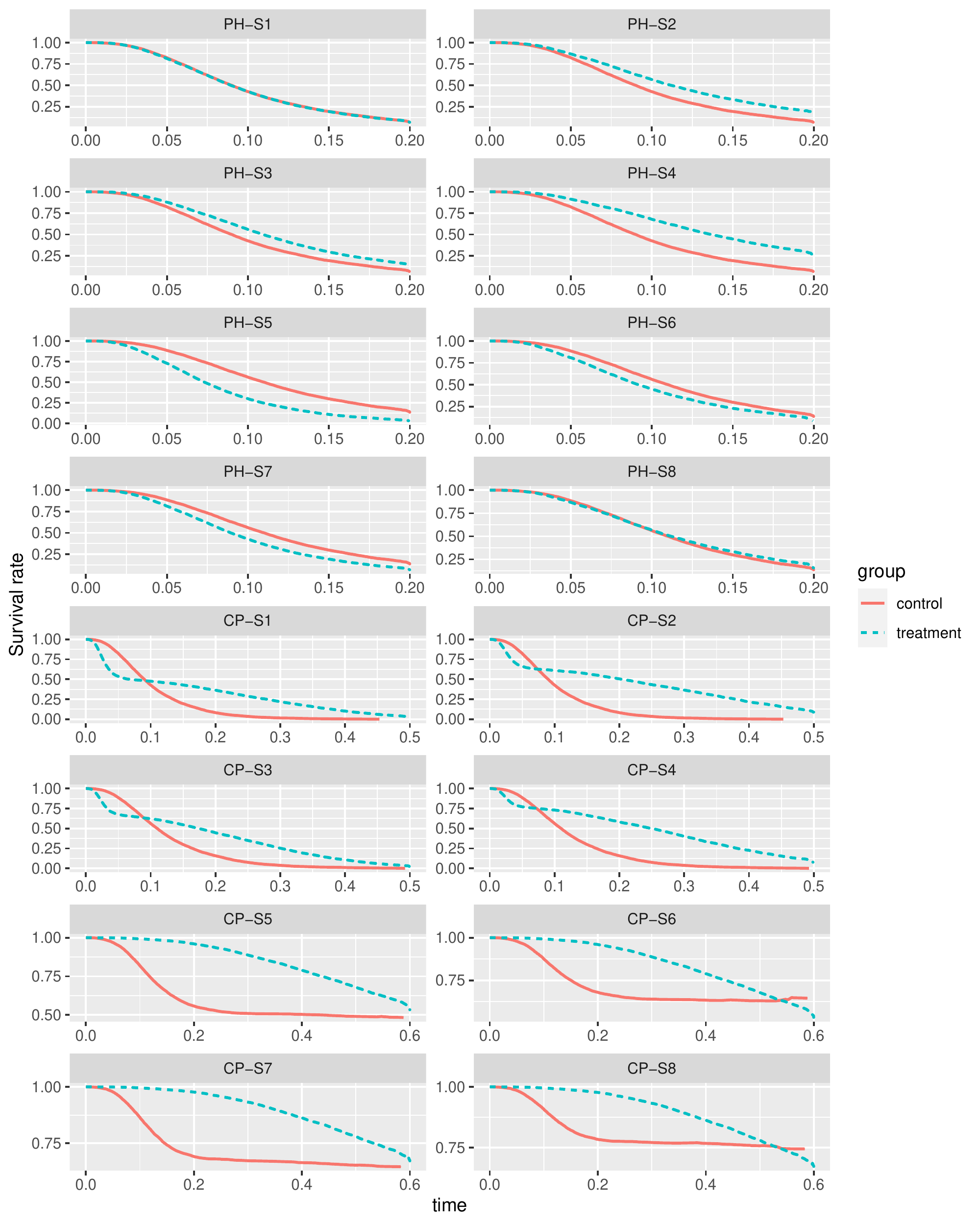}
  \caption{Survival curves of simulation scenarios}
  \label{sim_fig1}
  \end{center}
\end{figure}

For CP-S1--CP-S8, we use the regression model with CPs to generate the simulation data.
Let the survival times $T$ be given by
\begin{align}
    \log T = \beta_0 + \beta_1 x_1 + \beta_2 x_2 + \beta_3 x_1 x_3 + \log(\varepsilon), \,\,\,\,\,
    x_3 = \left \{ \begin{array}{cc}
        a & (x_2 \le 50\mbox{ }\mathrm{ (change \, point)}) \\
        b & (x_2 > 50\mbox{ }\mathrm{ (change \, point)})
    \end{array}
    \right.
\end{align}
where $\beta_3$ is a covariate parameter with CP = $50$.
The parameter settings are presented in Table \ref{sim_Table1} and the related survival curves are shown in Figure \ref{sim_fig1}.
We generate 10,000 datasets with $n=10, 20, \dots, 90, 100$ subjects for each scenario. We determine the power to detect the differences in RMST between groups without adjusting for covariates, known as the naive approach. For Andersen's and Tian's methods, the statistical model for detecting group differences is as follows:
\begin{align}
    \exp(\ga_1+x_1\ga_2+x_2\ga_3).
    \label{eq:Sim_Cox}
\end{align}
When the p-value of $\gah_2$ for the null hypothesis $H_0:$ $\ga_2=0$ is less than $0.05$, we consider the group differences to be detected. For reference, we also determine the power of Cox's PH model to detect differences between the groups. The parametric function of Cox's PH model is given by (\ref{eq:Sim_Cox}).
\begin{table}[H]
  \begin{center}
\caption{Parameter settings of simulation scenarios \label{sim_Table1}}
\begin{tabular}{|c|c|c|c|c|c|cc|c|c|cc|}
\hline
 & & & & & & \multicolumn{2}{c|}{True mean of $x_2$} & & & \multicolumn{2}{c|}{Percentage of censored data} \\
Scenario & $\tau$ & $\beta_0$ & $\beta_1$ & $\beta_2$ & $\beta_3$ & \multicolumn{1}{c|}{Treatment} & Control & $a$ & $b$  & \multicolumn{1}{c|}{Treatment} & Control \\ \hline
PH-S1 & 0.2    & 2 & 0 & -0.1 & - & \multicolumn{1}{c|}{50}& 50 &  -   &    -   & \multicolumn{1}{c|}{0.1} & 0.1 \\
PH-S2 & 0.2    & 2 & 0 & -0.1 & - & \multicolumn{1}{c|}{50}& 50 &  -   &    -   & \multicolumn{1}{c|}{\textcolor{red}{\bf 0.4}} & 0.1 \\
PH-S3 & 0.2    & 2 & \textcolor{red}{\bf 0.2} & -0.1 & - & \multicolumn{1}{c|}{50}& 50 &  -   &    -   & \multicolumn{1}{c|}{0.1} & 0.1 \\
PH-S4 & 0.2    & 2 & \textcolor{red}{\bf 0.2} & -0.1 & - & \multicolumn{1}{c|}{50}& 50 &  -   &    -   & \multicolumn{1}{c|}{\textcolor{red}{\bf 0.4}} & 0.1 \\
PH-S5 & 0.2    & 2 & 0 & -0.1 & - & \multicolumn{1}{c|}{\textcolor{red}{\bf 48}}& \textcolor{red}{\bf 52} &  -   &    -   & \multicolumn{1}{c|}{0.1} & 0.1 \\
PH-S6 & 0.2    & 2 & 0 & -0.1 & - & \multicolumn{1}{c|}{\textcolor{red}{\bf 48}}& \textcolor{red}{\bf 52} &  -   &    -   & \multicolumn{1}{c|}{\textcolor{red}{\bf 0.4}} & 0.1 \\
PH-S7 & 0.2    & 2 & \textcolor{red}{\bf 0.2} & -0.1 & - & \multicolumn{1}{c|}{\textcolor{red}{\bf 48}}& \textcolor{red}{\bf 52} &  -   &    -   & \multicolumn{1}{c|}{0.1} & 0.1 \\
PH-S8 & 0.2    & 2 & \textcolor{red}{\bf 0.2} & -0.1 & - & \multicolumn{1}{c|}{\textcolor{red}{\bf 48}}& \textcolor{red}{\bf 52} &  -   &    -   & \multicolumn{1}{c|}{\textcolor{red}{\bf 0.4}} & 0.1 \\\hdashline
CP-S1  & 0.5    & 2 & -0.5      & -0.1      & 2 & \multicolumn{1}{c|}{50}& 50      & 1   & -0.5  & \multicolumn{1}{c|}{0.1}       & 0.1     \\
CP-S2  & 0.5    & 2 & -0.5      & -0.1      & 2 & \multicolumn{1}{c|}{50}& 50      & 1   & -0.5  & \multicolumn{1}{c|}{\textcolor{red}{\bf 0.4}}       & 0.1     \\
CP-S3  & 0.5    & 2 & -0.5      & -0.1      & 2 & \multicolumn{1}{c|}{\textcolor{red}{\bf 48}}& \textcolor{red}{\bf 52}      & 1   & -0.5  & \multicolumn{1}{c|}{0.1}       & 0.1     \\
CP-S4  & 0.5    & 2 & -0.5      & -0.1      & 2 & \multicolumn{1}{c|}{\textcolor{red}{\bf 48}}& \textcolor{red}{\bf 52}      & 1   & -0.5  & \multicolumn{1}{c|}{\textcolor{red}{\bf 0.4}}       & 0.1     \\\hdashline
CP-S5  & 0.6    & 4 & -0.5      & -0.1      & 3 & \multicolumn{1}{c|}{50}& 50      & 1   & -0.5  & \multicolumn{1}{c|}{0.1}       & 0.1     \\
CP-S6  & 0.6    & 4 & -0.5      & -0.1      & 3 & \multicolumn{1}{c|}{50}& 50      & 1   & -0.5  & \multicolumn{1}{c|}{\textcolor{red}{\bf 0.4}}       & 0.1     \\
CP-S7  & 0.6    & 4 & -0.5      & -0.1      & 3 & \multicolumn{1}{c|}{\textcolor{red}{\bf 48}}& \textcolor{red}{\bf 52}      & 1   & -0.5  & \multicolumn{1}{c|}{0.1}       & 0.1     \\
CP-S8  & 0.6    & 4 & -0.5      & -0.1      & 3 & \multicolumn{1}{c|}{\textcolor{red}{\bf 48}}& \textcolor{red}{\bf 52}      & 1   & -0.5  & \multicolumn{1}{c|}{\textcolor{red}{\bf 0.4}}       & 0.1  \\\hline   
\end{tabular}
\\
      \footnotesize{For PH-S2--PH-S8, changes from PH-S1 are shown in red. For CP-S2--CP-S4, changes from CP-S1 are shown in red. For CP-S6--CP-S8, changes from CP-S5 are shown in red.}  \end{center}
\end{table}

\subsection{Simulation results}
\begin{figure}[H]
  \begin{center}
  \includegraphics[width=15cm]{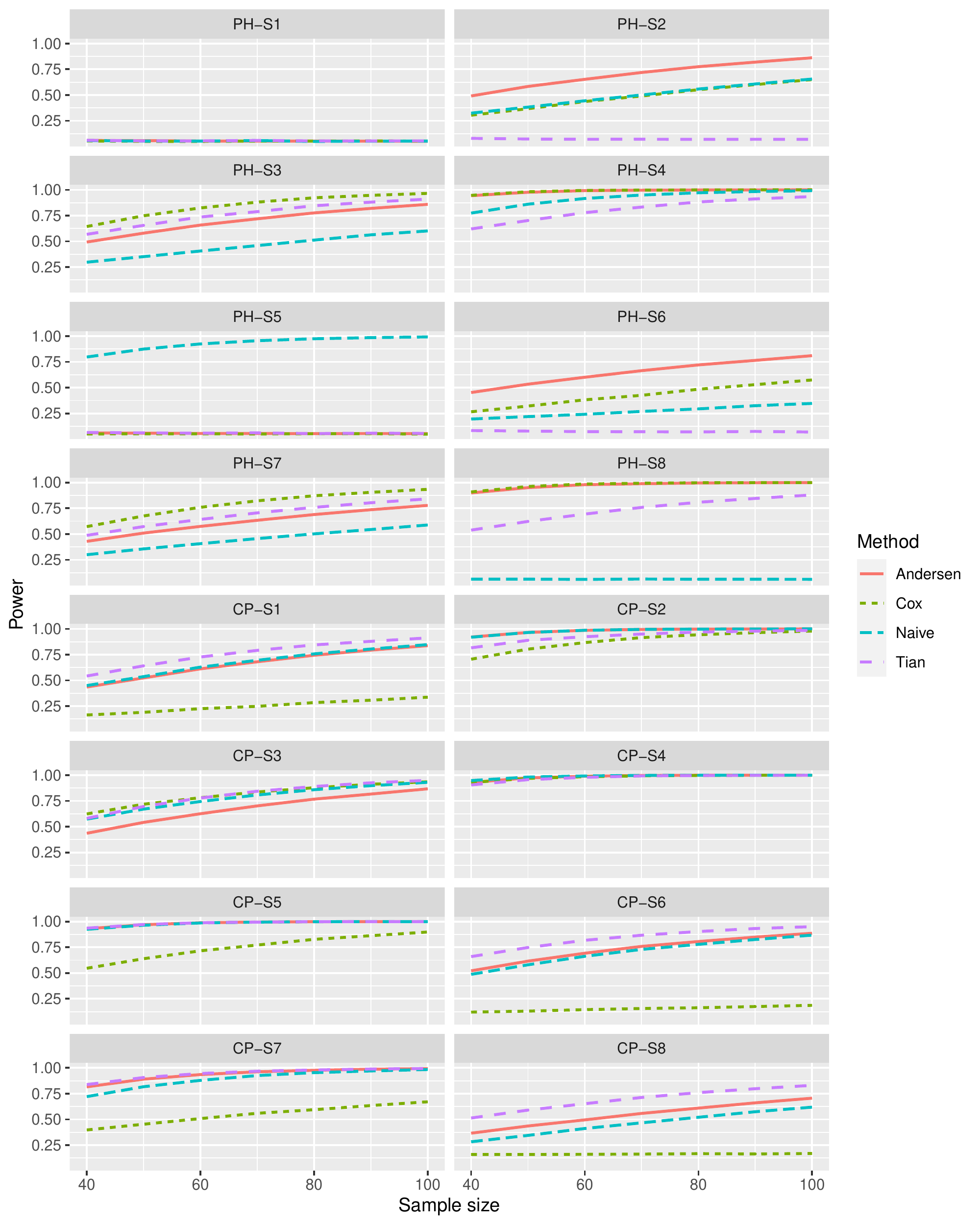}
  \caption{Power of simulation scenarios}
  \label{sim_fig2}
  \end{center}
\end{figure}

We present the results of the 16 simulation scenarios in Figure \ref{sim_fig2}. For PH-S1, the type-1 error rate can be controlled at the nominal significance level by all methods. For PH-S2, Tian can control the type-1 error rate. However, the type-1 error rates of Andersen, Cox and naive are inflated as the number of subjects increases. For PH-S3, the powers of Andersen, Tian, and Cox adjusting for covariates are higher than naive. For PH-S4, Andersen, Cox and naive have the highest power but Tian had slightly lower power. For PH-S5, the Andersen, Tian, and Cox models give a nominal significance level, whereas the naive approach produces an inflated type-1 error rate as the number of subjects increases because of the lack of adjustment in the covariate. For PH-S6, Tian can control the type 1 error rate but other methods inflated the type-1 error rate as the number of subjects increases. For PH-S7, the powers of Andersen, Tian, and Cox adjusting for covariates are higher than naive. For the PH-S8, Andersen and Cox have the highest power, Tian had slightly lower power, and naive does not increase in power even when the number of subjects was increased. 

For the CP scenarios, the power of all methods tends to rise as the number of subjects increases. Notably, all methods have very high power in CP-S4, and similarly for CP-S5, except for Cox's model, in which the power increase more slowly.

\section{Re-analysis of Data from Two Clinical Trials}
To evaluate Andersen's and Tian's methods, we re-analyzed the data of two actual clinical trials using the RMSTREG PROCEDURE in SAS. The SAS analysis program code is included in the Supplemental Material.

\subsection{Re-analysis of clinical trial data in primary biliary cholangitis}
\begin{figure}[H]
  \begin{center}
  \includegraphics[width=15cm]{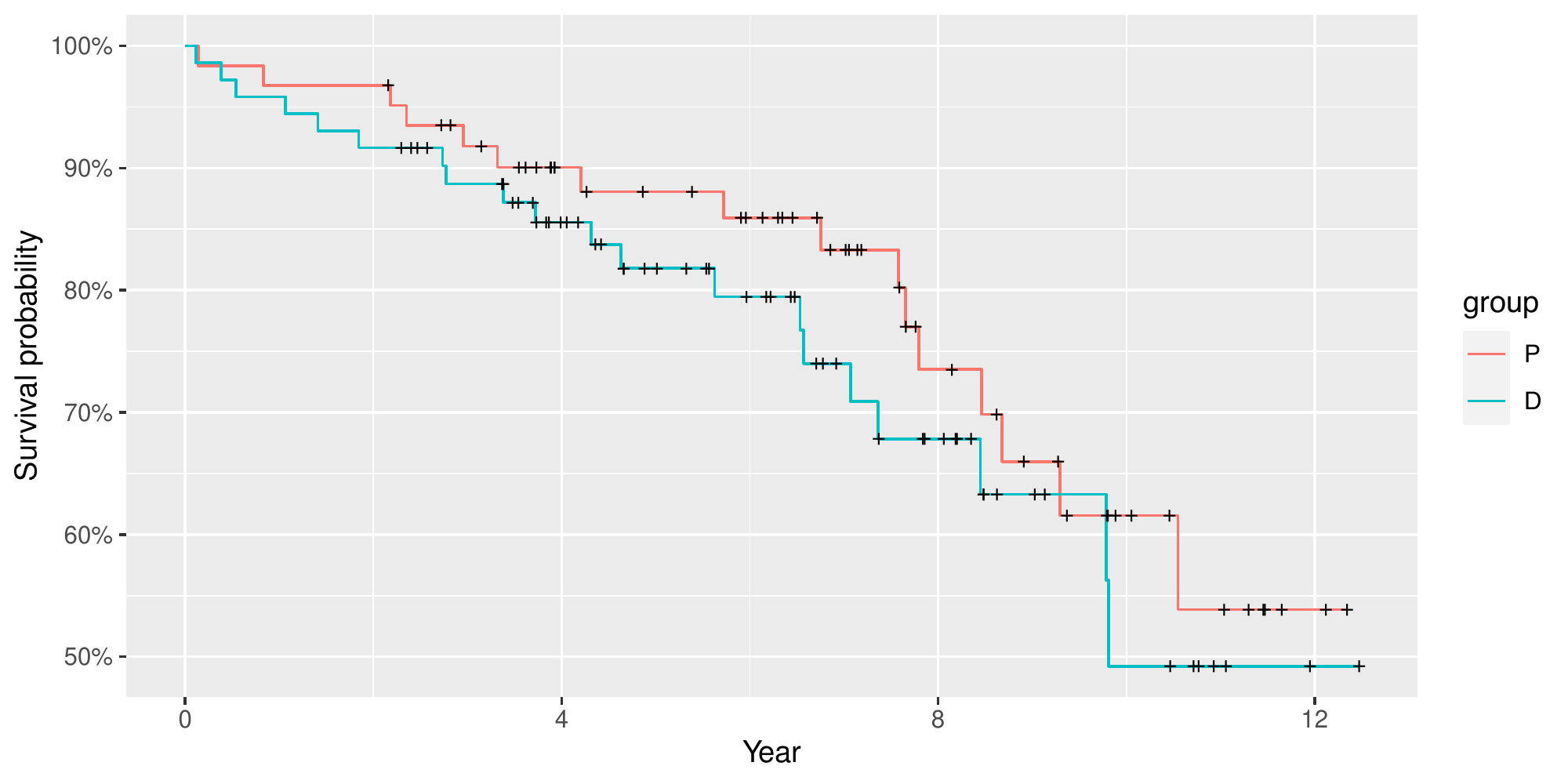}
  \caption{Survival curves of PBC data}
  \label{MK_Figure1}
        \footnotesize{P: Placebo group, D: D-penicillamine group, + denotes censored.}
  \end{center}
\end{figure}

We evaluated the performance of Andersen's and Tian's methods using the primary biliary cholangitis (PBC) dataset in the survival package of R~\cite{therneau2015}. This dataset comes from a randomized placebo-controlled trial of primary biliary cholangitis conducted between 1974 and 1984. The original PBC data include non-randomized patients. Considering only randomized patients, we selected 134 patients (72: D-penicillamine group, 62: placebo group) without hepatomegaly or enlarged livers. The percentage of censoring in the D-penicillamine group is 72.22\% (52/72) and the percentage of censoring in the placebo group is 74.19\% (46/62). The survival curves are shown in Figure \ref{MK_Figure1}. 
The covariates are edema (0: no edema, 0.5: untreated or successfully treated, 1: edema despite diuretic therapy), serum bilirubin (mg/dl), serum albumin (mg/dl), standardized blood clotting time (protime), and age (years). The restricted time was 12.34 yrs. The link function is set to identity to understand the interpretation of the analysis results easily.

In the re-analysis results, the naive RMST $\pm$ standard error is $8.88\pm 0.62$ in the placebo group and $7.96\pm 0.61$ in the D-penicillamine group, with a $-0.92\pm 0.54$ difference between the two groups (D-penicillamine group $-$ placebo group). The confidence interval (p-value) is $[-1.97, 0.10]$ $(0.086$). The difference in RMST between groups adjusted by Andersen's method is $-0.56\pm 0.67$ $[-1.88, 0.76]$  $(0.405)$, and that between groups adjusted by Tian's method is $-0.93\pm 0.43$ $[-1.78, -0.09]$  $(0.030)$. Table \ref{MK_Table1} presents the estimators of all model parameters. The survival curves for the PST are shown in Figure \ref{MK_Figure3} in the appendix. We can confirm that the survival curves of the original data and the pseudo-data are different. 

\begin{table}[H]
  \begin{center}
\caption{Analysis results of Andersen's and Tian's methods in PBC data\label{MK_Table1}}
\begin{tabular}{|l|l|l|l|l|l|l|}\hline
& \multicolumn{3}{c|}{Andersen}& \multicolumn{3}{c|}{Tian}\\\hline
Parameter & Mean & SE & [CI] (p-value) & Mean & SE & [CI] (p-value)\\ \hline
Intercept &20.20&6.49 & [7.47, 32.93] (0.002)&-3.39 &3.70 &[-10.63, 3.86] (0.360) \\
Treatment group &-0.56 &0.67 &[-1.88, 0.76] (0.405) &-0.93 &0.43 &[-1.78, -0.09] (0.030) \\
Edema (1) &-1.29 &2.58 &[-6.35, 3.77] (0.618) &-5.70 &1.70 &[-9.04, -2.35] (0.001)\\
Edema (0.5) &0.39 &1.68 &[-2.90, 3.68] (0.814) &-4.40 &0.52 &[-5.42, -3.38] (0.000)\\
Bilirubin &-0.49 &0.14 &[-0.76, -0.23] (0.000) &-0.11 &0.11 &[-0.32, 0.11] (0.332)\\
Albumin &0.24 &1.08 &[-1.87, 2.35] (0.823) &0.89 &0.57 &[-0.24, 2.01] (0.122)\\
Protime &-0.48 &0.51 &[-1.48, 0.53] (0.351) &0.50 &0.23 &[0.04, 0.96] (0.033)\\
Age &-0.11 &0.03 &[-0.17, -0.04] (0.002) &0.05 &0.02 &[0.00, 0.09] (0.056)\\\hline
\end{tabular}
\\
      \footnotesize{SE: Standard Error. CI: $95\%$ Confidence Interval. Treatment group and edema are fixed effects.}
  \end{center}
\end{table}

\subsection{Re-analysis of clinical trial data in malignant glioma}
\begin{figure}[H]
  \begin{center}
  \includegraphics[width=15cm]{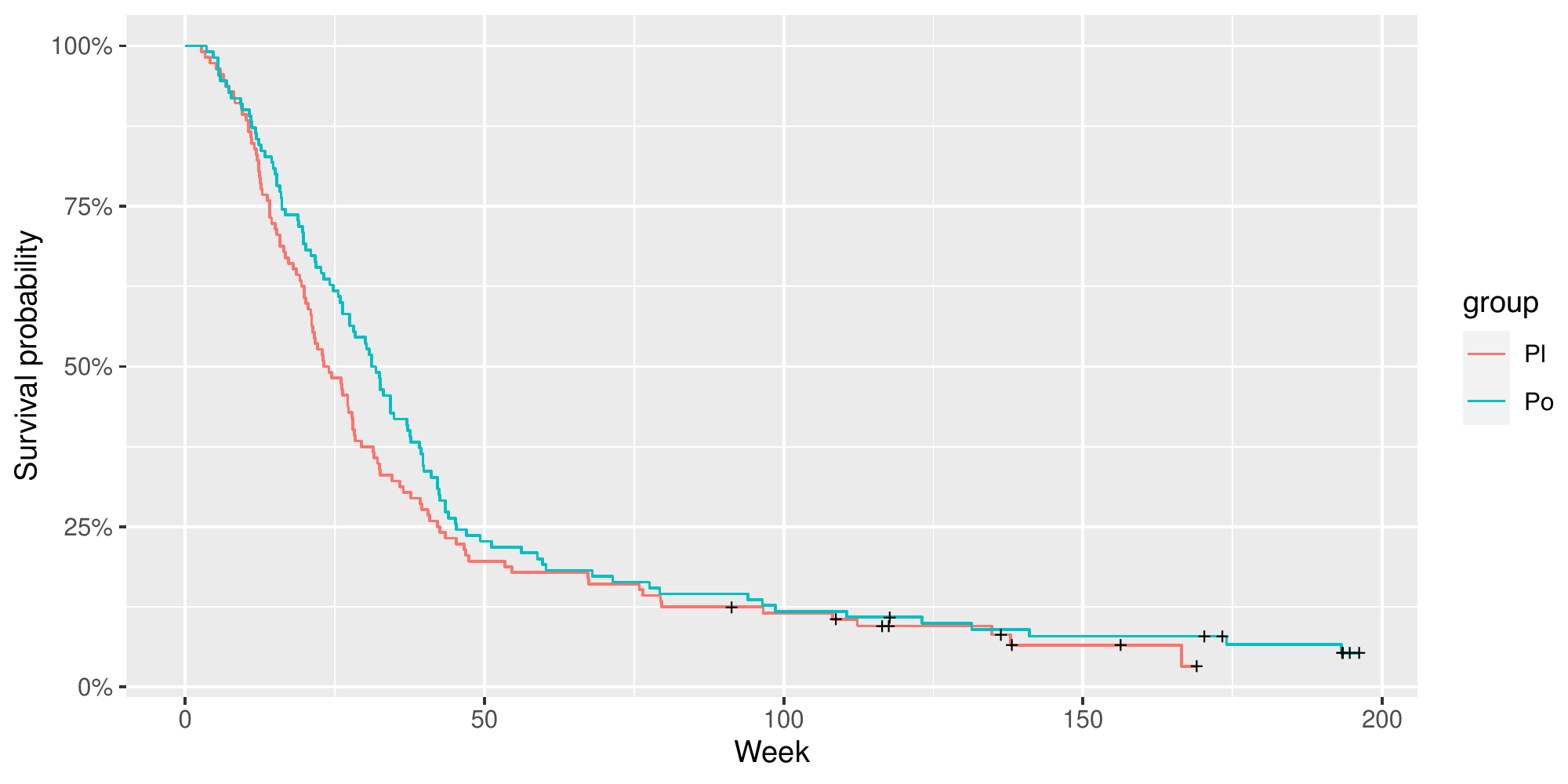}
  \caption{Survival curves of malignant glioma data}
  \label{MK_Figure2}
        \footnotesize{Pl: Placebo group, Po: Polymer group, + denotes censored.}
  \end{center}
\end{figure}

We also evaluated the performance of Andersen's and Tian's methods using the randomized placebo-controlled clinical trial data for malignant glioma reported by Brem et al.~\cite{brem1995}. The analysis data include 110 patients in the treatment group and 112 patients in the placebo group. The percentage of censoring in the treatment group is 6.36\% (7/110) and the percentage of censoring in the placebo group is 7.14\% (8/112). The survival curves are shown in Figure \ref{MK_Figure2}.
The covariates are age (years), years from diagnosis, Karnofsky performance score (0: < 70, 1: $\leq$ 70), race (0: other, 1: white), sex (0: female, 1: male), nitro (0: no, 1: yes), tumor histopathology at implantation (path) (1: glioblastoma, 2: anaplastic astrocytoma, 3: oligodendroglioma, 4: other), and grade (0: quiescent, 1: active). The restricted time point was 150 week. The link function is set to identity to understand the interpretation of the analysis results easily.

The naive RMST $\pm$ standard error is $40.51\pm 2.03$ in the placebo group and $47.40\pm 2.18$ in the polymer group, with a $6.88\pm 6.29$ difference between the two groups (polymer group $-$ placebo group). The confidence interval (p-value) is $[-5.45, 19.21]$ $(0.274$). The difference in RMST between groups adjusted by Andersen's method is $-0.56\pm 0.67$ $[-1.88, 0.76]$  $(0.405)$ and that between groups adjusted by Tian's method is $-0.93\pm 0.43$ $[-1.78, -0.09]$  $(0.030)$. Table \ref{MK_Table2} gives the estimators of all model parameters. The survival curves for the PST are shown in Figure \ref{MK_Figure4} in the appendix. We can confirm that the survival curves of the original data and the pseudo-data are similar.

\begin{table}[H]
  \begin{center}
\caption{Analysis results of Andersen's and Tian's methods in malignant glioma data\label{MK_Table2}}
\begin{tabular}{|l|l|l|l|l|l|l|}\hline
& \multicolumn{3}{c|}{Andersen}& \multicolumn{3}{c|}{Tian}\\\hline
Parameter & Mean & SE & [CI] (p-value) & Mean & SE & [CI] (p-value)\\ \hline
Intercept&75.12&21.12&[33.74, 116.51] (0.000)&47.18&20.57&[6.87, 87.49] (0.022)\\
Treatment group&7.41&5.77&[-3.89, 18.71] (0.199)&12.10&5.33&[1.64, 22.55] (0.023)\\
Age&-0.57&0.23&[-1.01, -0.13] (0.012)&-0.13&0.19&[-0.49, 0.24] (0.499)\\
Dx&1.70&1.48&[-1.21, 4.61] (0.252)&1.71&1.10&[-0.44, 3.87] (0.120)\\
PS&12.69&5.25&[2.41, 22.97] (0.016)&8.75&5.41&[-1.85, 19.36] (0.106)\\
Race&-17.54&13.46&[-43.93, 8.84] (0.193)&-8.47&9.76&[-27.61, 10.66] (0.385)\\
Sex&11.60&5.35&[1.12, 22.08] (0.030)&-0.29&5.2&[-10.48, 9.9] (0.955)\\
Nitro&-9.98&5.50&[-20.76, 0.8] (0.070)&-15.6&4.01&[-23.47, -7.73] (0.000)\\
Path (4)&47.74&20.07&[8.4, 87.07] (0.017)&22.46&11.49&[-0.06, 44.98] (0.051)\\
Path (3)&37.19&13.23&[11.26, 63.12] (0.005)&22.45&13.69&[-4.37, 49.28] (0.101)\\
Path (2)&19.30&9.22&[1.22, 37.37] (0.036)&15.14&7.57&[0.31, 29.98] (0.045)\\
Grade&-15.68&11.74&[-38.7, 7.34] (0.182)&-9.23&9.05&[-26.96, 8.5] (0.307)\\\hline
\end{tabular}
\\
      \footnotesize{SE: Standard Error. CI: $95\%$ Confidence Interval. Treatment group and edema are fixed effects.}
  \end{center}
\end{table}

\section{Discussion}
We evaluated the performance of Andersen's and Tian's baseline covariate adjustment methods because we could not decide which method to include in a statistical analysis plan for a clinical trial. Andersen's method calculates the PST using a leave-one-out approach for all subjects, including censored subjects, and performs regression analysis using all pseudo-values. Tian's method uses IPCW to estimate the parameters of the regression model, and also considers censoring. Therefore, there is a significant difference between the two methods in the process of adjusting the covariates. In Andersen's method, the interpretation of pseudo-values is difficult. We consider Tian's method to be more intuitive than Andersen's method. In this paper, we confirmed the intuitive behavior of Andersen’s and Tian’s methods using a simple example dataset.

Through a simulation study, we confirmed that Tian's method has higher power than Andersen's method. Tian's method is able to control the type-1 error rate independently of censor bias, whereas Andersen's method and Cox's PH model inflate the type-1 error rate. For scenarios PH-S3 and PH-S6, Cox's PH model only has slightly higher power than the RMST. Royston and Parmar~\cite{royston2013}, Tian et al.~\cite{tian2018}, and Huang and Kuan~\cite{huang2018} also found that the log-rank test had slightly higher power than the RMST under the PH assumption. In the CP scenarios, where the PH assumption does not hold, the RMST is more powerful. In PH-S4, PH-S8, CP-S2, and CP-S4, when the treatment group has a higher proportion of censored data, Andersen's method and Cox's PH model exhibit the highest power. However, because the censors are biased toward the treatment group, Andersen's method and Cox's PH model give an inflated type-1 error rate. Hence, we consider Tian's method to be more appropriate in terms of retaining statistical power in certain situations.

We re-analyzed clinical PBC trial data provided by the R survival package and malignant glioma trial data provided by Piantadosi \cite{piantadosi2017}. For the PBC data, Andersen's method implies that bilirubin and age affect the RMST. In contrast, Tian's method implies that the treatment group, edema (1, 0.5), and protime affect the RMST. These differences in results can be attributed to differences in methodology. The percentage of censoring in the placebo group is 74.19\% (46/62), and the percentage of censoring in the D-penicillamine group is 72.22\% (52/72). Thus, the high incidence of censoring may have contributed to the difference in results. For Andersen's method, there is no difference in bilirubin between groups and an age difference of 2.9 (years) (47.4 in the placebo group and 50.3 in the D-penicillamine group). Because the D-penicillamine group is older and the estimate of the age parameter is negative, the difference is smaller than the adjustment for covariates. For Tian's method, there is no difference in protime between groups. In the placebo group, edema (0) is 87.5\% (=14/16) and edema (1) is 12.5\% (=2/16), while in the D-penicillamine group, edema (0) is 75\% (=15/20), edema (0.5) is 10\% (=2/20), and edema (1) is 15\% (=3/20). Although there is a small difference in the edema distribution between groups, little adjustment is made for differences in RMST because of the small number of observed events. We discuss the adjusted difference in RMSTs in more detail in the appendix. 

For the malignant glioma data, Andersen's method implies that age, PS, sex, and path (4, 3, and 2) affect the RMST. Tian's method implies that the treatment group, nitro, and path (4 and 2) affect the RMST. The reason for the difference in estimation results is the different estimation methods. For Andersen's method, the difference in RMST is almost the same because there is no significant difference in the background information. For Tian's method, there is no difference in nitro between groups; for the placebo group, path (1) is 68.3\%, path (2) is 14.4\%, path (3) is 16.3\%, and path (4) is 1.0\%, while in the polymer group, path (1) is 72.8\%, path (2) is 12.6\%, path (3) is 9.7\%, and path (4) is 4.9\%. The proportion of path (2) and path (3) in the polymer group is lower than in the placebo group. The estimates of the path (2) and path (3) parameters are positive, which widens the difference in RMST between the groups. We discuss the adjusted difference in RMSTs in more detail in the appendix.

In summary, the use of PSTs makes Andersen's method difficult to interpret, while Tian's method retains its power in various situations. Even when group differences occur with respect to the censor incidence rates and covariates, Tian's method maintains a nominal significance level for the type-1 error rate. Using actual trial data, we have confirmed that the PSTs are different from the actual data in the PBC case because many data are censored. Andersen's method should be used when the purpose is to adjust for group differences in RMST using the background of all subjects, including the censored subjects, rather than to adjust the backgrounds of only those subjects for whom the event occurred. We conclude that Tian's method can be recommended.

\bigskip
\noindent
{\bf Author Contributions.}
KH developed the method and prepared the simulation program code, JM reviewed the overall study design, and MK wrote the bulk of the paper. MK also analyzed the example data, simulation datasets, and clinical trial data.

\bigskip
\noindent
{\bf Acknowledgements.}
MK would like to thank Associate Professor Hisashi Noma for his encouragement and helpful suggestions.

\bibliography{main.bib} 
\bibliographystyle{unsrt} 

\appendix
\section{Appendix}\label{App}
\subsection{Generalized estimating equations of Andersen's method}
{\bf [Case where the link function is the identity function]}

The linear model is expressed as
\begin{align}
    \muh_i=\bbe^T\Z_i
\end{align}

\begin{align}
    \U_{PV}(\bbe) 
    &= \sum_{i=1}^n \U_i(\bbe) \nonumber \\
    &= \sum_{i=1}^n \left( \{\muh_i - \bbe^T \Z_i\}\Z_i  \right) \nonumber \\
    &= 0.
\end{align}

The estimator $\bbeh$ can be obtained as
\begin{align}
    \bbeh=\sum_{i=1}^n\muh_i\left(\sum_{j=1}^n\Z_j\Z_j^T\right)^{-1}\Z_i.
\end{align}

\noindent
{\bf [Case where the link function is the log function]}

The log-linear model is expressed as
\begin{align}
    \log(\muh_i)=\bbe^T\Z_i
\end{align}

The estimator $\bbeh$ can be obtained from the following equation. The equations below are not in closed-form and need to be estimated using numerical analysis.
\begin{align}
    \U_{PV}(\bbe) 
    &= \sum_{i=1}^n \U_i(\bbe) \nonumber \\
    &= \sum_{i=1}^n \left( \{\muh_i - \exp(\bbe^T \Z_i)\}\exp(\bbe^T \Z_i)\Z_i  \right) \nonumber \\
    &= 0.
\end{align}

\subsection{Estimating equations of Tian's method}
{\bf [Case where the link function is the identity function]}

The linear model is expressed as
\begin{align}
    X_i=\bbe^T\Z_i
\end{align}

\begin{align}
    \U_{IPCW}(\bbe) =\sum_{i=1}^n \frac{I(X_i \le C_i)\{ X_i - \bbe^T \Z_i\}}{\hat{S}(X_i)} \Z_i=0.
\end{align}

The estimator $\bbeh$ can be obtained as
\begin{align}
    \bbeh=\sum_{i=1}^n\frac{I(X_i \le C_i)\ X_i}{\hat{S}(X_i)}\left(\sum_{j=1}^n\frac{I(X_j \le C_j)}{\hat{S}(X_j)}\Z_j\Z_j^T\right)^{-1}\Z_i.
\end{align}

\noindent
{\bf [Case where the link function is the log function]}

The log-linear model is expressed as
\begin{align}
    \log(X_i)=\bbe^T\Z_i.
\end{align}

The estimator $\bbeh$ can be obtained from the following equation. This equation is not in closed-form and needs to be estimated using numerical analysis.
\begin{align}
    \U_{IPCW}(\bbe) =\sum_{i=1}^n \frac{I(X_i \le C_i)\{ X_i - \exp(\bbe^T \Z_i)\}}{\hat{S}(X_i)} \Z_i=0.
\end{align}

\subsection{Additional analyses of modified example dataset}
We considered two additional cases in which the age data were changed to confirm the influence caused by age. In the first case, there is no difference in average age between groups for all patients, including those censored, but there is a large difference in average age between groups of patients for whom events occur. The second case considers a difference in average age for all patients, including those censored. 

In the first case, the average age of patients for whom the event occurred is 75 for the treatment group and 67 for the control group. Detailed data are presented in Table \ref{MK_Table5}. The red font indicates changed data.

\begin{table}[H]
  \begin{center}
\caption{Example dataset\label{MK_Table5}}
\begin{tabular}{|c|c|c|c|c|c|c|c|c|c|}\hline
\multicolumn{5}{|c|}{Treatment group}& \multicolumn{5}{c|}{Control group}\\\hline
ID&ST & Censor & Age & PST &ID& ST & Censor & Age & PST\\ \hline
1&20 & Yes & 60 & 78.4 &7& 20 & No & \textcolor{red}{{\bf 60}} & 20.0\\
2&40 & No & 80 & 30.4 &8& 30 & Yes & \textcolor{red}{{\bf 70}} & 78.4\\
3&60 & Yes & 70 & 100.4 &9& 40 & No & 60 & 30.4\\
4&80 & No & 70 & 75.4 &10& 50 & No & 80 & 42.9\\
5&100 & Yes & 60 & 106.6 &11& 80 & Yes & 70 & 106.6\\
6&100 & Yes & 60 & 106.6 &12& 100 & Yes & 60 & 106.6\\\hline
\end{tabular}
\\
      \footnotesize{ST: Survival Time, PST: Pseudo-Survival Time}
  \end{center}
\end{table}

The analysis results are given in Table \ref{MK_Table6}. For Tian's method, the difference in RMST is increased by adjusting for the age because of the higher average age in the treatment group.

\begin{table}[H]
  \begin{center}
\caption{Differences in RMSTs of example dataset\label{MK_Table6}}
\begin{tabular}{|c|c|c|c|c|c|}\hline
Parameter & Naive & Andersen & Tian & Andersen (age) & Tian (age)\\\hline
Diff & 29.1 & 18.8 & 25.0 & 18.8 & \textcolor{red}{{\bf 28.1}}\\
Age & - & - & - & \textcolor{red}{{\bf -1.2}} & \textcolor{red}{{\bf -1.7}}\\\hline
\end{tabular}
\\
      \footnotesize{Andersen/Tian: statistical model includes only the treatment group as a fixed effect. Andersen (age)/Tian (age): statistical model includes the treatment group as a fixed effect and age as a covariate.}
  \end{center}
\end{table}

In the second case, the average age of all patients is 67 for the treatment group and 68 for the control group. Detailed data are presented in Table \ref{MK_Table7}. The red font indicates the changed data.

\begin{table}[H]
  \begin{center}
\caption{Example dataset\label{MK_Table7}}
\begin{tabular}{|c|c|c|c|c|c|c|c|c|c|}\hline
\multicolumn{5}{|c|}{Treatment group}& \multicolumn{5}{c|}{Control group}\\\hline
ID&ST & Censor & Age & PST &ID& ST & Censor & Age & PST\\ \hline
1&20 & Yes & 60 & 78.4 &7& 20 & No & 70 & 20.0\\
2&40 & No & 80 & 30.4 &8& 30 & Yes & \textcolor{red}{{\bf 70}} & 78.4\\
3&60 & Yes & 70 & 100.4 &9& 40 & No & 60 & 30.4\\
4&80 & No & 70 & 75.4 &10& 50 & No & 80 & 42.9\\
5&100 & Yes & 60 & 106.6 &11& 80 & Yes & 70 & 106.6\\
6&100 & Yes & 60 & 106.6 &12& 100 & Yes & 60 & 106.6\\\hline
\end{tabular}
\\
      \footnotesize{ST: Survival Time, PST: Pseudo-Survival Time}
  \end{center}
\end{table}

The analysis results are given in Table \ref{MK_Table8}. For Andersen's method, the difference in RMST is reduced by adjusting for age because of the higher average age in the control group.

\begin{table}[H]
  \begin{center}
\caption{Differences in RMSTs of example dataset\label{MK_Table8}}
\begin{tabular}{|c|c|c|c|c|c|}\hline
Parameter & Naive & Andersen & Tian & Andersen (age) & Tian (age)\\\hline
Diff & 29.1 & 18.8 & 25.0 & \textcolor{red}{{\bf 15.4}} & 24.5\\
Age & - & - & - & \textcolor{red}{{\bf -2.0}} & -2.3\\\hline
\end{tabular}
\\
      \footnotesize{Andersen/Tian: statistical model includes only the treatment group as a fixed effect. Andersen (age)/Tian (age): statistical model includes the treatment group as a fixed effect and age as a covariate.}
  \end{center}
\end{table}

\subsection{Survival curves of PSTs in two actual clinical trials}
\begin{figure}[H]
  \begin{center}
  \includegraphics[width=15cm]{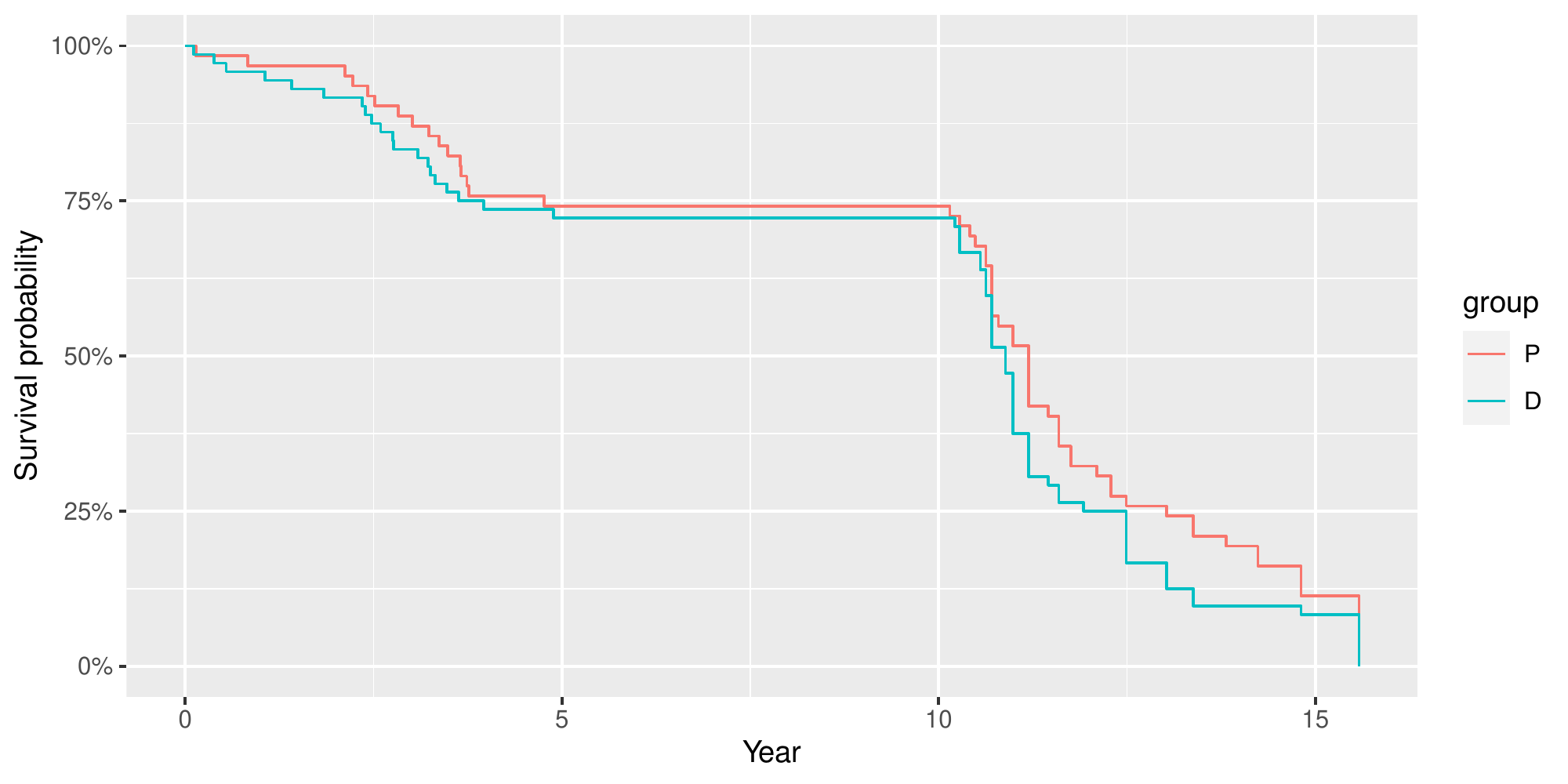}
  \caption{Survival curves of pseudo-values of PBC clinical trial data}
  \label{MK_Figure3}
  \end{center}
\end{figure}

The survival curves of the pseudo-values for PBC clinical trial data are shown in Figure \ref{MK_Figure3}. 

\begin{figure}[H]
  \begin{center}
  \includegraphics[width=15cm]{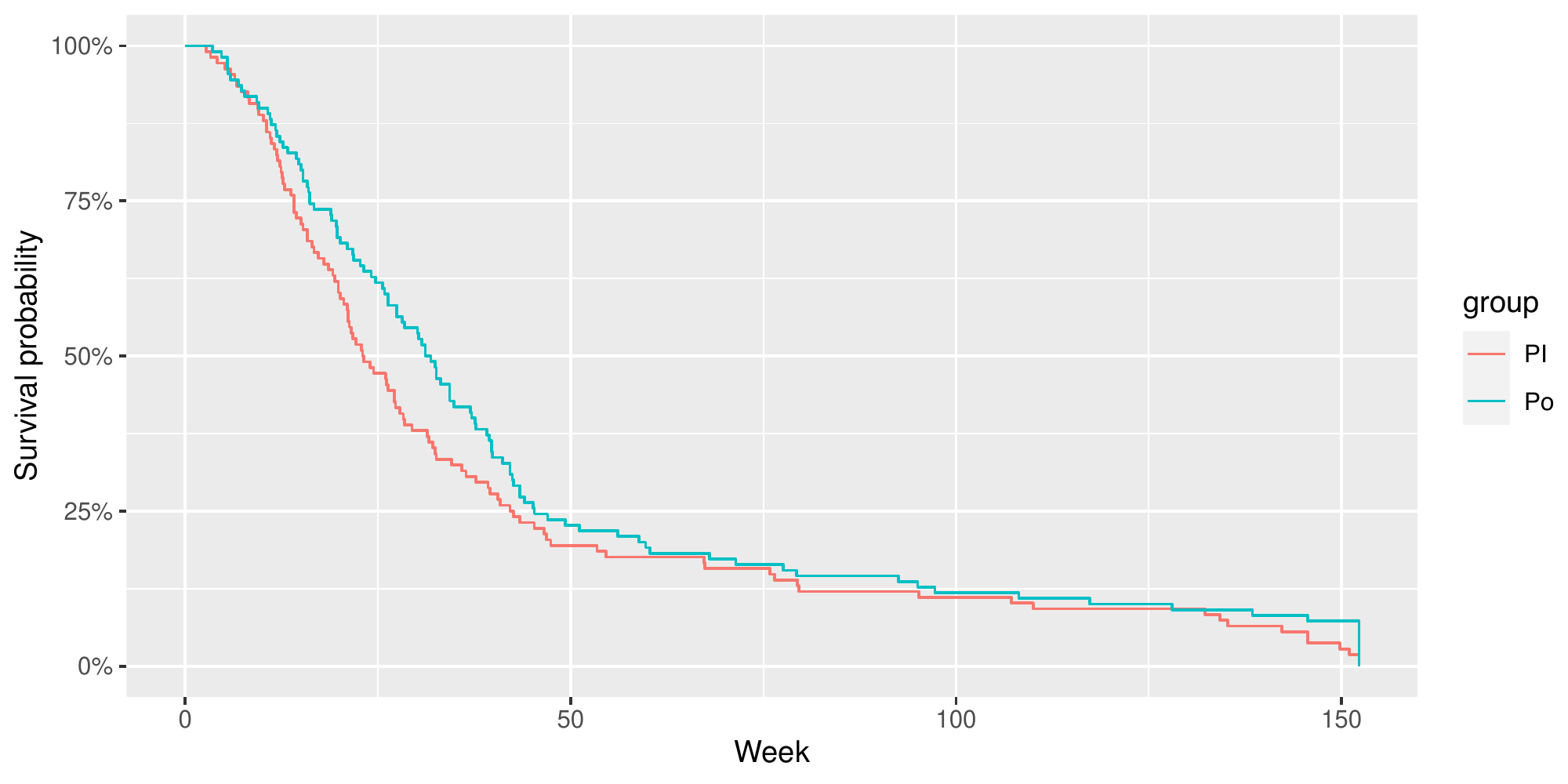}
  \caption{Survival curve of pseudo-values of malignant glioma clinical trial data}
  \label{MK_Figure4}
  \end{center}
\end{figure}

The survival curves of PSTs of malignant glioma clinical trial data are shown in Figure \ref{MK_Figure4}.

\subsection{Detailed analysis of data from two actual clinical trials}
For the PBC data, Andersen's method implied that bilirubin and age affect the RMST, whereas Tian's method implied that the treatment group, edema (1, 0.5), and protime affect the RMST. These differences in results were attributed to differences in methodology. The percentage of censoring in the placebo group is 74.19\% (46/62) and the percentage of censoring in the D-penicillamine group is 72.22\% (52/72). Thus, the high incidence of censoring may have contributed to the difference in results. For Andersen's method, the difference between groups for bilirubin and age, for which the confidence intervals did not include 0, was examined for all subjects, including the censored subjects. Bilirubin was measured at 2.1 in the placebo group and 1.9 in the D-penicillamine group, a difference of 0.2. The average age was 47.4 in the placebo group and 50.3 in the D-penicillamine group, a difference of 2.9. Because the D-penicillamine group was older and the estimate of the age parameter was negative, the difference was smaller than the adjustment for covariates. For Tian's method, the difference between groups for protime and the distribution of edema, for which the confidence intervals did not include 0, was examined for all subjects for whom the event occurred. Protime was equal to 11.1 in both the placebo and D-penicillamine groups. For the distribution of edema, edema (0) was 87.5\% (=14/16) and edema (1) was 12.5\% (=2/16) in the placebo group, and edema (0) was 75\% (=15/20), edema (0.5) was 10\% (=2/20), and edema (1) was 15\% (=3/20) in the D-penicillamine group. Although there is a small difference in the edema distribution between groups, little adjustment was made for differences in RMST because of the small number of observed events. 

For the malignant glioma data, Andersen's method implied that age, PS, sex, and path (4, 3, and 2) affect the RMST, whereas Tian's method implied that the treatment group, nitro, and path (4 and 2) affect the RMST. The reason for the difference in estimation results was the different estimation methods. For Andersen's method, the difference between groups for age, PS, sex, and path, for which the confidence intervals did not include 0, was examined for all subjects, including the censored subjects. The average age was 47.6 in the placebo group and 48.1 in the polymer group, a difference of 0.5. PS (1) was 50.0\% in the placebo group and 55.5\% in the polymer group, a difference of 5.5\%. The proportion of males was 61.6\% in the placebo group and 67.3\% in the polymer group, a difference of 5.7\%. In the placebo group, Path (1) was 65.2\%, path (2) was 14.3\%, path (3) was 17.9\%, and path (4) was 2.7\%, whereas in the polymer group, path (1) was 69.1\%, path (2) was 12.7\%, path (3) was 13.6\%, and path (4) was 4.5\%. The difference in RMST was almost the same because there was no significant difference in the background information. For Tian's method, the difference between groups for nitro and the proportion of path (2), for which the confidence intervals did not include 0, was examined for all subjects for whom the event occurred. Nitro was measured at 45.2\% in the placebo group and 49.5\% in the polymer group, a difference of 4.3\%. For the placebo group, path (1) was 68.3\%, path (2) was 14.4\%, path (3) was 16.3\%, and path (4) was 1.0\%, whereas in the polymer group, path (1) was 72.8\%, path (2) was 12.6\%, path (3) was 9.7\%, and path (4) was 4.9\%. The proportions of path (2) and path (3) in the polymer group were lower than in the placebo group. The estimates of the path (2) and path (3) parameters were positive, which widened the difference in RMST between the groups.

\end{document}